\documentstyle[aps,prl,epsfig,twocolumn]{revtex}

\begin{document}
\draft
\wideabs{
\title{Four-wave mixing in degenerate atomic gases}
\author{P. Villain, P. \"Ohberg, L. Santos, A. Sanpera and M. Lewenstein}
\address{Institut f\"ur Theoretische Physik, Universit\"at
Hannover,D-30167
Hannover, Germany}

\date{\today}

\maketitle

\begin{abstract}

We study the process of four-wave mixing (4WM) in ultracold 
degenerate atomic gases. In particular we address the problem of 4WM
in boson-fermion mixtures. We develop
an approximate description of such processes using 
asymptotic analysis of high order
perturbation theory taking into account quantum statistics. 
We perform also numerical simulations of 4WM in boson-fermion mixtures
and obtain an analytic and numerical estimate of the  efficiency 
of the process.

\end{abstract}

}

In the recent years atom optics has become a flourishing subject. With the 
successful experiments on Bose-Einstein condensation \cite{Cor95,Ket95,Hul95}
new possibilities to study linear\cite{Hann99}
and nonlinear atom-optics\cite{Meystre} 
with macroscopic wave packets have emerged. 
So far, the most spectacular experiment in  nonlinear atom optics
concerns the observation of  the so called 
four-wave mixing of bosonic matter waves\cite{Deng}. 
In this nonlinear process, three macroscopic matter wave packets interact 
and produce a fourth one. 
Also recently, Jin {\it et al.} \cite{Jin} 
have managed to trap and cool 
a sample of spin polarized $^{40}$K below the Fermi temperature in a harmonic
trap. This has triggered an outburst of activities in trying to understand 
the properties of ultracold fermionic systems\cite{fermion},
and has stimulated a great interest in the
studies of fermion-boson mixtures\cite{Moelmer}.
In particular, the question ``is non-linear atom optics with fermions
possible?'' has been posed\cite{phillips}.

In this Letter we investigate the process of four-wave mixing (4WM) in 
fermion-boson mixtures and demonstrate that under an appropriate choice of 
parameters it is possible to create a macroscopic fourth wave of fermions.
The recent experiment of Deng {\it et al.} \cite{Deng} with 
Bose-Einstein condensates have demonstrated that 4WM in
bosonic gases is a truly  macroscopic process  with an efficiency 
of the order of $6\%$. In this experiment Bragg pulses \cite{NIST} 
were used to create 
three condensate clouds, which then interacted through collisions. 
Four wave mixing of matter waves can also be described -- using the language
of optics -- as Bragg scattering from a grating. In this picture, 
two counter propagating condensates create the grating from which the
third condensate scatters, generating a fourth one. 

\begin{figure}

\centerline{\scalebox{0.3}{\epsffile{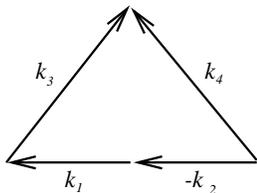}}}
\caption{In the reference frame in which $\vec k_1$ and $\vec k_2$ are 
collinear the scattering process
can always be described in a plane. The criterion $\vec k_4 =\vec k_1-\vec k_2+\vec k_3$ together 
with the conservation of energy $|\vec k_4|=|\vec k_3|$ gives the 
Bragg condition ${k_3}_x =k_1$ with $\vec k_1=-\vec k_2$ chosen in the 
direction of the x-axis.}   
\label{braggfig}
\end{figure}

An intuitive understanding of  Bragg scattering  for bosons 
may be obtained considering a homogeneous condensate in a box of 
volume $V$. For such a case, the Hamiltonian has the form:
\begin{equation}
\hat H = \sum_{\vec k} \epsilon_{\vec k} a^{\dag}_{\vec k} a_{\vec k} + 
{1\over 2}u_0\sum_{\vec k,\vec k',\vec q} a^{\dag}_{\vec k +\vec q} 
a^{\dag}_{\vec k'-\vec q} a_{\vec k'} a_{\vec k},\label{ham}
\end{equation}
where $u_0=\frac{4\pi\hbar^2 a}{mV}$ is the interaction strength 
proportional to the $s$-wave scattering length $a$ and $\epsilon_{\vec k}$ is 
the kinetic energy. 
Let us consider as an initial state 
$|i\rangle = |N_1,\vec k_1; N_2,\vec k_2; N_3,\vec k_3\rangle$
representing $N_i$ particles of momentum $\vec k_i$, $i=1,2,3$.
Assume for the moment 
that only one particle is scattered leading to the final state 
$|f\rangle = |N_1',\vec k_1;N_2',\vec k_2;N_3',\vec k_3;1,\vec k_4\rangle$ 
with $\vec k_4 \ne \vec k_{i}$ for $i=1,2,3$. Among all the 
processes which conserve momentum and energy the one corresponding to  
\begin{equation}
\vec k_4=\vec k_1 -\vec k_2 +\vec k_3 \label{bragg}
\end{equation}
is particularly favorable and represents in fact the first order 
Bragg scattering \cite{TrippOp,Tripp} (see Fig. \ref{braggfig}). 
The above process  
corresponds to the term $a_{\vec k_4}^\dagger a_{\vec k_2}^\dagger 
a_{\vec k_1} a_{\vec k_3}$ in the Hamiltonian. This term accounts for
the creation of a particle 
in the ``grating state'' $\vec k_2$ which is already 
macroscopically occupied. This leads to 
the appearance of a bosonic enhancement factor $\sqrt{N_2+1}$ in the 
transition amplitude associated with the process 
$|i\rangle \rightarrow |f\rangle$. From this point of view, due to
bosonic
enhancement Bragg scattering is the most probable process.

Let us extend the previous analysis to the case where many particles are 
scattered. Clearly, the most probable final state is of the form
$|f\rangle=|N_1',\vec k_1;N_2', \vec k_2;N_3',\vec k_3;N_4,\vec k_4\rangle$, 
where  $\vec k_4$  is given by 
Eq.(\ref{bragg}). This final state must respect the conservation of
momentum
\begin{equation}
\sum_{i=1}^3 N_i \vec k_i=
\sum_{i=1}^3 N_i' \vec k_i+N_4 \vec k_4 \, ,\label{c1}
\end{equation}
and the conservation of the particle number
\begin{equation}
\sum_{i=1}^3 N_i =\sum_{i=1}^3 N_i' +N_4\,. \label{c2}
\end{equation}
Eqs.(\ref{bragg}),(\ref{c1}) and (\ref{c2}) imply  conservation of
energy. In the reference frame in which $\vec k_1$ and $\vec k_2$ are 
collinear the scattering process is planar. Thus we have  five equations 
with six unknown parameters $N'_i,N_4,\vec k_4$. 
The actual 
value of $N_4$ can be determined by maximizing the transition probability 
$P_{if}$. Using perturbation theory with respect to the
off-diagonal elements of the Hamiltonian in Eq.\ref{ham}, we calculate the 
transition amplitude $T_{if}$. For $N_1=N_2=N_3$ this quantity 
exhibits a divergence for $\eta=N_4/N_{tot}=1/6\approx 16.7\%$, 
where $N_{tot}$ is the total number of particles. 
This divergence appears due to the simplicity of the model and can be 
removed by using wave packets instead of plane waves. In any case,
$P_{if}$ will be strongly peaked at $ N_4\sim  N_{tot}/6$.
We expect that the actual efficiency of the process -- which in
principle has to be calculated non perturbatively -- will 
be of the same order of magnitude. 
Furthermore, our analysis predicts a saturation of the efficiency 
with increasing  number of atoms, which 
can be understood as a consequence of the Bose statistics. These
results are in agreement with the experimental results of Ref. \cite{Deng}.

We turn now to the case of fermions. The previous analysis clearly stresses 
the role of bosonic enhancement; the macroscopic occupation of the
states  that form the
grating select Bragg scattering as the most favorable process.
From this point of view, the
use of a fermionic grating would lead to a rather poor Bragg
scattering. For this reason we consider Bragg scattering of  
an incoming cloud of fermions on a bosonic grating.
In our model, fermions and bosons interact via 
two-body $s$-wave scattering. 
The incoming fermions must obey 
momentum and energy conservation as in the case of boson-boson scattering.
For fermions there will be inevitably  a momentum spread due to 
Fermi statistics.
In order to fulfill the Bragg condition, illustrated in Fig. \ref{braggfig},
the momentum spread in the fermionic 
sample $\Delta k$, must be restricted to $\Delta k \ll 
|\vec k_1 -\vec k_2|$.
This condition can be achieved by trapping the fermions initially 
in a sufficiently shallow trap which results in a well 
localized momentum distribution. 

Solving the complete many-body scattering problem for the fermions is a 
formidable task. In our case we can simplify the situation assuming first
that the fermions do not interact. This is a good approximation 
as long as we  consider a polarized Fermi gas. 
The interaction of fermions of only one specie  is 
entirely due to $p$-wave scattering which is negligible at very low 
temperatures \cite{Jin2}. Second, we model the bosonic grating by a
scalar potential proportional to the local condensate density.
In doing this we sacrifice in some sense the effects of  the statistics. 
In fact, the Fermi statistics appears in our model only 
through the preparation of the initial
state, which must obey Pauli principle.  
We also neglect any back-action from the fermions on the bosons. This 
last approximation is valid in a situation where the boson density is much 
larger than the fermion density. 
Finally, since the dynamics is basically 
two dimensional (see Fig. 1)  we restrict our numerical simulations to 2D. 
With these assumptions we solve the Schr\"odinger equation for the
fermions with 
an effective external potential of the form:
\begin{equation}
V_f(x,y)=u_{BF} n_b(x,y)\,,\label{gpot}
\end{equation}
with $u_{BF}={{8\pi\hbar^2 a}\over {mL}}$. Here $a$ denotes the 
fermion--boson $s$-wave 
scattering length, $m$ the (same) mass for both fermions and bosons, 
$L$ the thickness of the cloud in $z$-direction, whereas
$n_b(x,y)$  the equilibrium density of the condensate.
For the bosons, the trap potential is a combination of a harmonic
potential in the {\it y}-direction and a periodic 
structure in {\it x}-direction
created for instance by a standing-wave laser,
\begin{equation}
V_b(x,y)={1\over2}m\Omega_c^2 y^2+U_{l}\cos^2(|\vec k_1-\vec k_2| x).
\label{vbos}
\end{equation} 
We have solved 
numerically the  Gross-Pitaevskii equation 
with the potential in Eq. \ref{vbos}, 
in order to determine the equilibrium bosonic density $n_b(x,y)$,
which, in
turn, results in a periodic potential $V_f(x,y)$ for the fermions.
The back action of fermions on bosons can be neglected provided 
$u_{BF}n_f(x,y)/\mu \ll 1$ where $\mu$ is the 
chemical potential of the trapped condensate and $n_f(x,y)$ 
the fermion density. In our simulation this ratio was kept
smaller than 0.01.

Our procedure is now as follows: initially  the fermions are trapped in a 
2D potential of the form $V(x,y)= m \Omega^2 (x^2+y^2)$ centered at
$(x_0,y_0)$. The number of fermions is such
that the Fermi level still fulfills the
condition $k_F\ll |\vec k_1 -\vec k_2|$.
The trap is then removed and a momentum kick 
$\hbar {\vec k_3}$ is given to the fermions by 
illuminating the sample with a detuned laser propagating 
in the direction of $\vec k_3$ \cite{Deng,MIT99} so the fermion
cloud scatters from the grating. We finally  monitor the density
of fermions to obtain the efficiency of the process. 

The wave functions of the non interacting fermions fulfill all the same
Schr\"odinger equation
\begin{equation}
i\hbar{\partial\over{\partial t}}\Psi_i=
\left[ -{{\hbar^2}\over{2m}}\nabla_i^2+V_f(x,y)
\right]\Psi_i, i=1,...,N_3,
\end{equation}
but with different initial conditions for each fermion. The initial
states are the eigenstates of the displaced harmonic potential used 
for trapping the fermions,
\begin{equation}
\Psi_{n_x,n_y}(x,y)=A_{n_x,n_y} e^{i\vec k_3 \cdot \vec x} 
e^{-{1\over 2} (\tilde x^2+\tilde y^2)} H_{n_x}(\tilde x)
H_{n_y}(\tilde y), 
\label{eig}
\end{equation}
where $\tilde x=(x-x_0)/\sqrt{\hbar/m\Omega}$ and 
$\tilde y=(y-y_0)/\sqrt{\hbar/m\Omega}$, $H_{n_x}$ denotes the  
Hermite polynomials and $A_{n_x,n_y}$ is 
the normalization constant.

The scattering of the fermions can now be numerically simulated one by 
one. We monitor the density in the region
where the Bragg scattered fermions are supposed to appear. 
In our model, for the fermion cloud we use $^{40}$K atoms 
and a trap frequency of $\Omega/2\pi=10$ Hz while for the bosonic trap 
we use a frequency of $\Omega_c=40\Omega$. This produces a
narrow grating compared to the size of the fermionic cloud. 
Furthermore, we have taken the  
scattering length value as $a=6.0$ nm and $L=30$ $\mu$m with $N_c=2\cdot 
10^5$ bosons
in the condensate.
The grating has an associate  wave number of $k_1=1.3$ $\mu$m$^{-1}$ 
($\vec k_1=-\vec k_2$) and $k_F/k_1<0.7$. The initial fermion cloud is 
positioned at $x_0=y_0=-40$ $\mu$m and the momentum kick is settled 
to $\vec k_3=(1.3,1.3)\mu$m$^{-1}$.

\begin{figure}

\centerline{\scalebox{0.3}{\epsffile{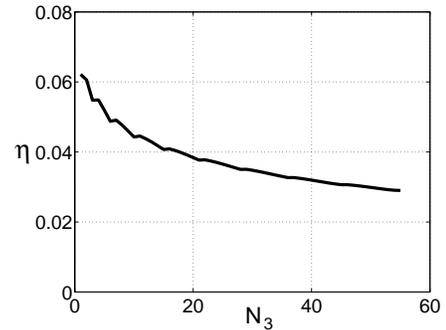}}}
\caption{Numerical estimate of the efficiency $\eta$ of the 4WM 
versus the number of incoming fermions $N_3$.  
$\eta$ decreases  for larger values of $N_3$ due to the spread of the 
momentum for the incoming fermions.} 
\label{eff}
\end{figure}

\begin{figure}

\centerline{\scalebox{0.4}{\epsffile{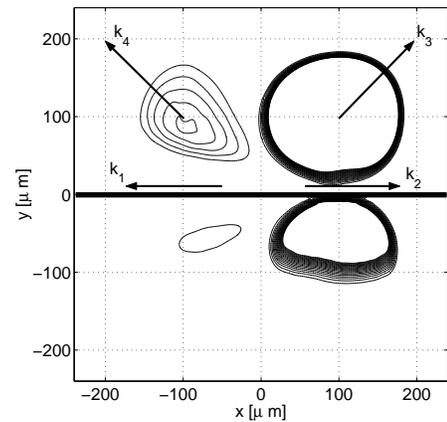}}}
\caption{A Snapshot of the density of the fermionic cloud 
after the scattering. In this case 
$3\%$ of the atoms are found in the Bragg direction. The grating is 
represented by the two counter propagating waves $\vec k_1=-\vec k_2$}   
\label{den}
\end{figure}

In Fig.\ref{eff} we show the 
efficiency $\eta=N_4/N_3$ of the Bragg process as a function of the total 
number of fermions $N_3$. 
With an increasing number of particles ($N_3$) the 
efficiency decreases due to the increase of the momentum spread of
the fermions.
Fig. \ref{den} displays a 
snapshot of the cloud after the scattering. 
We observe that approximately 3\% of the cloud is scattered in the Bragg 
direction.
In this figure one can also see that a part of the fermionic cloud is 
reflected due to the chosen relation between the incoming fermion kinetic 
energy and the contrast together with the maximum of the grating potential.
Note also the appearance of the reflected wave packet 
in the direction  $-\vec k_3$ which in fact corresponds to a Bragg 
reflection. Reflections are drastically reduced 
if the scattering length is negative since in this case the potential $V_f(x,y)$ becomes attractive.
The Bragg scattering relies on the periodicity and contrast
of the grating and it is present for both 
positive and negative scattering lengths. 
Also, for a fixed interaction time
the efficiency of the Bragg process  
decreases with the contrast of the grating.

In order to understand the above results we have developed an approach
similar to the one used for bosons, based on perturbative analysis
using  plane waves.
We want to study the 
efficiency of the process starting with the fermionic state 
$|i\rangle=|1,{\vec k_3}+{\vec \chi_1};...;1,{\vec k_3}+{\vec
\chi_{N_3}}\rangle$ 
and considering a generic final state 
in which $N_4$ holes are created in the Fermi sea of states with
momentum close to $\vec k_3$.
There are $j={N_3 \choose N_4}$
different orthogonal final states $|f_j\rangle$.
In order to simplify the problem we assume that the transition 
amplitude does not depend strongly on $|f_j\rangle$, which is 
a valid assumption as long as  $\Delta k\ll |\vec k_1 -\vec k_2|$. 
We can then
evaluate the probability $P(N_4)$ of 
scattering $N_4$ particles by taking an average value 
of the transition amplitude so that $P(N_4)\approx {N_3 \choose N_4}
{|\bar{T}_{if_j}|^2}$. 
The mean efficiency of the process defined as
$\eta=\langle N_4\rangle/N_3$ is calculated then using 
the previous probability distribution $P(N_4)$.
We stress that the aim of this calculation is not to obtain 
an exact expression for the efficiency, but rather to estimate 
whether the scattering process can change macroscopically a many 
particle fermionic state.

\begin{figure}

\centerline{\scalebox{0.4}{\epsffile{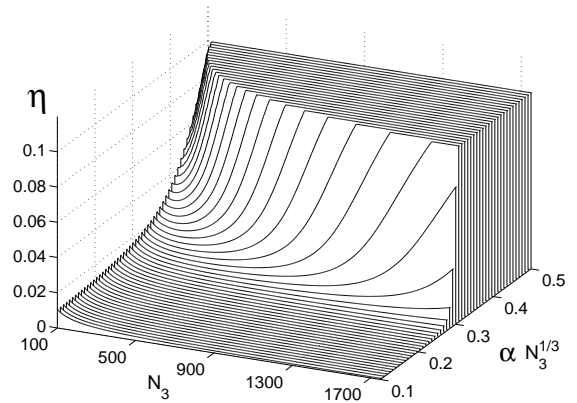}}}
\caption{Analytical estimate of the Bragg process efficiency. The 
model is limited by the 
assumption $\eta\ll 1$. For small values 
of $\alpha$ the efficiency decreases with increasing $N_3$.}
\label{pro}
\end{figure}

We begin by calculating the transition amplitude $T_{if_1}$ 
between the initial state $|i\rangle$ and the final state
$|f_1\rangle$ which corresponds to depletion of the first 
$N_4$ states of the Fermi sea around 
$\vec k_3$. The 
macroscopically populated states of the bosonic grating 
provide a bosonic enhancement factor  $\sqrt{(N_2+p)}$ $(p=1,2,...,N_4)$.
Using a time dependent perturbative method \cite{Schiff} one obtains
\begin{equation}
T_{if_1}\propto  U_{BF}^{N_4} \frac{\Gamma}
{(E_i-E_{N_4})...(E_i-E_1)} \label{amp}, 
\end{equation}
where the bosonic enhancement amounts to
\begin{equation}
\Gamma=\frac{\sqrt{N_1!}\sqrt{(N_2+N_4)!}}{\sqrt{(N_1-N_4)!} 
\sqrt{N_2!}}.
\end{equation}
The energies of the intermediate states (corresponding to the
scattering
of $p$-fermions) are given by
\begin{equation}
E_{p}=E_B+\frac{\hbar^2}{2m}\left\{\sum_{j=p+1}^{N_3}({\vec k_3}+
{\vec \chi_j})^2+\sum_{j=1}^{p}({\vec k_4}+\vec \chi_j)^2 \right\}
\end{equation}
with $U_{BF}=\frac{8\pi\hbar^2 a}{mV}$, $E_i=E_{p=0}$ and 
${\vec k_2}=k_{2}\hat e_x =-k_1 \hat e_x$. We denote by $E_B$ the  
energy of the bosons which is constant during the process.
This $T_{if_1}$ is the transition 
amplitude corresponding to one of the $N_4!$ possible paths going from 
$|i\rangle$ to $|f_1\rangle$. 
Hence there are $N_4!$ different $T_{if_1}$'s and we 
estimate their contribution by taking an average value. 
To this aim we consider the 
$\vec \chi$ 's as independent random variables distributed between 
$-k_F$ and $k_F$ with a flat probability distribution 
$\propto\theta(k_F-\chi)$. 
Neglecting correlations between different $\vec \chi$ 's is a valid 
approximation if $N_4\ll N_3$. 
Consequently we consider this regime in the following, and restrict
ourselves to the values of $N_4\leq 0.1 N_3$.
The resulting probability of having $N_4$ Bragg scattered fermions is 
\begin{equation}
P(N_4)\propto 
\left(\frac{1}{\alpha^2}\right)^{N_3-N_4}
\frac{1}{(N_3-N_4)!},
\end{equation}
where we have assumed $N_1=N_2$.
This Poissonian probability distribution depends on the number of 
fermions $N_3$ and on the parameter 
\begin{equation}
\alpha=\frac{2\sqrt{10}\,\pi\, a\, n_b}{k_2 k_F},
\end{equation} 
which is also $N_3$-dependent through the dependence  on the
Fermi momentum $k_F$. The 
parameter $\alpha$ contains also the boson density $n_b$.
In Fig. \ref{pro} we display the efficiency $\eta$, 
(defined as the 
mean value of $N_4$ over $N_3$) as a function of $N_3$ and the 
$N_3$--independent parameter $\alpha N_3^{1/3}$.
We observe that for small values of $\alpha$ the efficiency decreases 
with increasing $N_3$. On the contrary, with larger $\alpha$  
the efficiency can reach larger values, only limited by the
assumptions of the model. i.e $\eta\leq 0.1$.

In summary, we have discussed the effects of Bose and Fermi statistics on 
four-wave mixing processes. For a pure bosonic process the Bose 
statistics sets a fundamental limit in the efficiency. In the case of an 
incoming fermionic cloud, the result depends strongly on the values of the 
various physical parameters involved in the problem. 
On one hand, we have performed a numerical 
analysis which shows that 
the momentum spread $\Delta k \ll |\vec k_1-\vec k_2|$ is a crucial 
parameter in order to obtain a macroscopic fourth 
wave. On the other hand, the analytical treatment  
which assumes that Bragg condition is always fulfilled, exhibits  
an interplay between the statistical and collisional effects 
leading to a  decreasing efficiency for small $N_3$ and increasing
efficiency for large $N_3$. 
Typically, for small 
$\alpha$, the value of $\langle N_4 \rangle$ is negligible meaning a
small efficiency. On the contrary, for sufficiently big values of $\alpha$, 
the efficiency is only limited by the assumptions of the model, namely the 
lack of correlations between the scattered particles.
The efficiency can then reach
values of the order of a few percent which characterizes the creation of a
macroscopic fermionic fourth wave. 

The authors wish to thank W. Ketterle for helpful discussions.
We acknowledge support from Deutsche Forschungsgemeinschaft (SFB
407) and the TMR network ERBXTCT96-0002.

\end{document}